\documentclass[12pt]{article} \textheight 8.5in \textwidth 6.5in
\usepackage{amsmath,amsfonts,latexsym,amssymb}

\oddsidemargin 0in \topmargin -.35in \title{The Arrow of Time and the
Initial Conditions of the Universe} \author{Robert M. Wald\thanks{\tt
rmwa@midway.uchicago.edu} \\ \it Enrico Fermi Institute and Department
of Physics \\ \it University of Chicago \\ \it 5640 S.~Ellis Avenue,
Chicago, IL~60637, USA}
\begin{document}

\maketitle
\begin{abstract}

The existence of a thermodynamic arrow of time in the present universe
implies that the initial state of the observable portion of our
universe at (or near) the ``big bang'' must have been very
``special''. We argue that it is not plausible that these special
initial conditions have a dynamical origin.

\end{abstract}

There is no question that our present universe displays a
thermodynamic arrow of time: We all have observed phenomena in our
everyday lives where entropy is seen to increase significantly, but no
one has ever reliably reported an observation of entropy decrease in a
macroscopic system.  This fact raises a number of obvious questions.
Some of these questions have straightforward answers---although they
largely ``beg the question'':

\begin{itemize}

\item {\bf Question}: Why is there a thermodynamic arrow of time in
the present universe?

\item {\bf Answer}: Because the present entropy of the universe is
very low compared with how high it could be. It is well understood
that during the normal course of dynamical evolution, an isolated
system will spend the overwhelming majority of its time in a state of
(nearly) maximum entropy. Therefore, if we find a system to be in a
state of low entropy, it is overwhelmingly likely that we will observe
it to evolve to a state of higher entropy. The probability of
observing it to evolve to a state of lower entropy is vanishingly
small.

\end{itemize}

\begin{itemize}

\item {\bf Question}: Why is the entropy of the present universe low?

\item {\bf Answer}: Because in the past, the entropy was even lower
than it is now. Although the entropy has been increasing with time, it
has not had time to reach (or even come very close to) a state of
maximum entropy. This explains why it is still low today. In other
words, the entropy of the present universe is low because it was much
lower a billion years ago; it was very low a billion years ago because
it was even lower than that two billion years ago, etc.  Clearly, this
line of reasoning leads us back to (or, at least, very near to) the
``big bang'': The reason why the entropy of the present universe is
low is that the entropy of the universe at or near the ``big bang''
was extremely low \cite{penrose}.

\end{itemize}

The above claim that the entropy of the very early universe must have
been extremely low might appear to blatantly contradict the ``standard
model'' of cosmology: There is overwhelmingly strong reason to believe
that in the early universe, matter was (very nearly) uniformly
distributed and (very nearly) in thermal equilibrium at uniform
temperature. Doesn't this correspond to a state of (very nearly)
maximum entropy, not a state of low entropy? In fact, for a system
subject only to short-range forces as usually considered in textbooks,
in a state of maximum entropy, matter would be homogeneously
distributed and at uniform temperature. But, the situation changes
dramatically when an unscreened, long-range force such as gravity is
present. Even in Newtonian gravity, for a sufficiently large system,
the entropy can always be increased by clumping the system and using
the binding energy that is thereby released to heat up the system. In
Newtonian gravity, this phenomenon is usually referred to as ``Jeans'
instability'' or ``gravithermal instability''. For point particles in
Newtonian gravity, there is no bound to the binding energy, and an
isolated, self-gravitating system will simply become more and more
clumpy with time. General relativity effectively provides a cutoff to
this process via the formation of black holes. Nevertheless, in
general relativity, for a sufficiently large system, the state of
maximum entropy will not correspond to a homogeneous distribution of
matter but rather will contain a large black hole. The entropy of this
black hole (as given by the Bekenstein-Hawking formula) will be
enormously greater than a state at the same energy and volume where
matter is distributed homogeneously. In this way, it can be understood
that the early universe was in a state of extremely low entropy
compared with how high it's entropy could have been \cite{penrose}.

Before proceeding further, I should add some caveats to the above
discussion. The arguments that underlie our understanding of
statistical physics and thermodynamics are based upon having a time
translationally invariant system whose dynamics are ``ergodic'' to a
suitable degree. Such arguments certainly do not apply
straightforwardly to general relativistic systems. In particular,
although general relativistic systems are diffeomorphism covariant,
they are not ``time translation invariant'' in the sense required to
apply the usual arguments of statistical physics. Furthermore, in a
cosmological setting it seems clear that dynamics cannot, in any
sense, be ``ergodic'': In what sense could an open universe that
expands forever (or a closed universe that recollapses within finite
time) be said to ``sample'' a suitably large portion of its allowed phase
space? Finally, general relativity is a classical field theory and, as
such, would not be expected to have a sensible thermodynamics in any
case (for the same reason as classical electromagnetism suffers from
the ``ultraviolet catastrophe''); we should need to have a quantum
theory of gravity and a complete understanding of all of its
fundamental degrees of freedom before one could hope to obtain a full
understanding of the thermodynamic behavior of gravity. For all of the above
reasons, there does not presently exist a general notion of
``gravitational entropy'', and one should exercise considerable
caution when applying thermodynamic arguments to general relativistic
systems, such as the entire universe. In particular, recalling that
for a closed universe in general relativity there is no meaningful
notion of the ``total energy of the universe'', I see no reason to
expect that there will be a meaningful notion of the ``total entropy
of the universe''. Nevertheless, there is very strong encouragement
from all of the remarkable results obtained in black hole
thermodynamics (see, e.g., \cite{w1}) that in (quantum) general
relativity, some notion of entropy will exist and the basic form of
the laws of thermodynamics will survive. On account of this, I am
reasonably confident that the essential content of the assertions and
arguments of the preceding paragraphs will also survive in some form.

The answers to the first two questions above lead us to the following
question:

\begin{itemize}

\item {\bf Question}: What caused the very early universe to be in a
very low entropy state?

\end{itemize}

Here I do not have a simple answer to propose. But it would seem that,
logically, there are two basic ways to try to account for why the
initial state of the very early universe was so ``special'': (i) The
initial state of the universe was, in fact, ``completely
random''. However, dynamical evolutionary behavior was
then responsible for making (at least our portion of) the universe be
``very special''. (ii) The universe simply came into existence in a
very special state.

Viewpoint (i) appears to be presently favored by the overwhelming
majority of cosmologists. It is usually taken for granted that the
universe must have come into existence in a ``random state'', as
though there were a ``dartboard of initial conditions'', and the
actual initial conditions of our universe were selected by the throw
of an unskilled and blindfolded creator. Perhaps the best developed
and most popular of the ideas for producing a universe like the one
that we see from random initial conditions is chaotic inflation. Here,
one postulates the existence of a scalar field (the ``inflaton'') with
suitable properties. With random initial conditions for the metric and
scalar field, most portions of the universe should recollapse or
expand to emptyness on a timescale of the order of the Planck time. However,
there also should, by chance, exist regions of sufficiently large size
in which conditions are right for the onset of inflation. These
regions would then expand exponentially for many e-folding times, so
that they would dominate the volume of the universe. Within each
inflated region, the universe would be extremely homogeneous and
isotropic (and extremely spatially flat), with the only significant
deviations from homogeneity and isotropy being those produced by quantum
fluctuations. These quantum fluctuations would then result in observable
deviations from isotropy in the microwave background and provide the
seeds for the formation of the structure observed in the present
universe.

Inflationary models are extremely successful in predicting the kind of
deviations from homogeneity and anisotropy that we observe in our
universe. Indeed, it is quite difficult to come up with alternative
models that so naturally produce Gaussian fluctuations of the correct
amplitude and ``scale-free'' spectrum \cite{hw}. However, I do not
believe that inflationary models---or, for that matter, any other
dynamical mechanism---can provide a satisfactory answer to the above
question as to why (a suitably large portion of) the very early
universe was in such a ``special'' state, and I therefore also do not
believe that inflation can provide a satisfactory explanation for the
origin of the thermodynamic arrow of time\footnote{Recently, Carroll
and Chen \cite{cc} have proposed that ``spontaneous inflation'' can
account for a locally observed arrow of time in a universe that is
time symmetric on ultra-large scales. In their model, the universe has
entropy growing unboundedly in both the past and future. The universe
is ``normally'' (i.e., in most of the spacetime) a nearly empty
deSitter spacetime, but, occassionally, thermal fluctuations produce
regions of inflation that result in a large increase of entropy in
that region, and a corresponding locally observed arrow of time.  In
their model, episodes of inflation would not be favored over episodes
of ``deflation'' (i.e., eras of exponential contraction, in which the
entropy decreases); indeed, episodes of inflation would dominate in
the distant future, whereas episodes of deflation would dominate in
the distant past. I do not find their proposal to be plausible, but,
in essentially all other respects, the discussion of the issue of the
origin of the thermodynamic arrow of time given in \cite{cc} is
compatible with the viewpoints taken here.}.  My unhappiness with
attempts to use inflation or any other dynamical mechanism to try to
account for our observable universe being in very special state
despite it's having started in a random/generic state can be seen as
follows. In essence, in order to dynamically evolve from an assumed
``random'' initial state to the kind of very ``special'' state we
observe, it is necessary to invoke rare and/or highly unlikely events.
For example, in chaotic inflation, the initial conditions needed to
produce an inflating patch in the early universe are very ``special'';
most regions would not inflate and would not evolve to a universe that
looks anything like ours. Of course, it is true that, nevertheless,
some regions are bound to inflate.  Indeed, if the universe is
infinite, the probability of having an inflating patch (and, indeed,
infinitely many such patches) is $1$.  Thus, there is no difficulty in
arguing that it is {\it possible} that the portion of the universe
that we observe arose from an inflating patch as described in the
chaotic inflation scenario.  But in an infinite universe starting with
``random'' initial conditions, the probability of having a hugh patch
that directly evolves---without inflation---to a region
indistinguishable from the observed universe also is $1$ (as is the
probability of producing a universe indistinguishable from ours except
that all elephants wear pink dresses), so it also is {\it possible}
that the portion of the universe that we observe arose in this manner.
In order for the chaotic inflationary scenario or other dynamical
mechanisms to do better than this, it is necessary to argue that,
within the context of the model, observers in the universe are {\it
likely} to see a universe like the one we see; the presently observed
universe should not merely be a (highly unlikely) {\it possibility}
that is allowed in the model but rather should be a {\it prediction}
of the model.

Unfortunately, as I now shall argue, it appears inevitable that the
attempt to make predictions within the context of models of this sort
leads one down an essentially circular path, with little, if any,
possibility of attaining any explanatory power. In order to predict
what an observer should see, one must modulate the probabilities of
the various possible cosmological occurrences by the ``selection
effect'' that any observed portion of the universe must contain
conscious life in order to be observed. This modulation of
probabilities is usually referred to as the ``anthropic principle''.
Now, the probabilities of various cosmological occurences---such as
the probability that a given patch will inflate---are already
extremely difficult to estimate: We have, at best, only a vague notion
of what we mean by ``random'' or ``generic'' initial conditions; we
know very little about the true physical processes that may have
occurred in the very early universe; and, in any case, the
probabilities of rare occurrences are notoriously difficult to
estimate (since rare occurrences often do not arise in ``expected''
ways). However, our ignorance of the probabilities of various
cosmological occurrences is truly dwarfed by our (nearly) total
ignorance of the probability of the existence of observers, since we
know virtually nothing about what is really required to produce
conscious life. Therefore, it is usual practice in such arguments to
substitute the requirement that the observed portion of the universe
contain observers with the requirement that the observed portion of
the universe have some key features like ours, such as the presence of
stars and galaxies. (Obviously, in making such a substitution, one is
effectively assuming that the only way---or, at least, the most
probable way---of producing conscious life is by following a route
very similar to ours; I, personally, do not find this assumption to be
plausible.) But then, logically, the only ``prediction'' being made is
the determination of the probability of having a region of the
universe that is similar to the observed universe subject to the
constraint that this region possess certain key features that are
known to be present in the observed universe. Even if the calculation
of this probability could be reliably done, I fail to see what one
would learn from it. In particular, I fail to see in what sense it
would provide an ``explanation'' of why the observable universe is in
the state we find it to be in.

It seems to me to be far more plausible that the answer to the above
question as to why the very early universe was in a very low entropy
state is that it came into existence in a very special state. Of
course, this answer begs the question, since one would then want to
know why it came into existence in a very special state, i.e., what
principle or law governed its creation. I definitely do not
have an answer to this question. But I believe that it will be more
fruitful to seek an answer to this question than to attempt to pursue
dynamical explanations.

\medskip

This research was supported in part by NSF grant PHY00-90138 to the
University of Chicago.

\end{document}